\documentstyle[preprint,eqsecnum,aps,psfig]{revtex}
\newif\iftightenlines\tightenlinesfalse
\tightenlines\tightenlinestrue
\def\eslt{E\llap/_T}
\def\to{\rightarrow}
\def\Re{{\cal R \mskip-4mu \lower.1ex \hbox{\it e}}\,}
\def\Im{{\cal I \mskip-5mu \lower.1ex \hbox{\it m}}\,}

\def\te{\tilde e}

\def\ttau{\tilde \tau}
\def\tg{\tilde g}
\def\tmu{\tilde \mu}

\def\tnu{\tilde\nu}
\def\tell{\tilde\ell}
\def\tq{\tilde q}
\def\tw{\widetilde W}
\def\tz{\widetilde Z}
\def\D0{{\rm D}\emptyset}
\begin{document}
%
%%%%%%%%%%%%%%%%%%%%%%%%%%%%%%%%%%%%%%%%%%%%%%%%%%%%%%%%%%%%%%%%%
%
%%%%%%%%%%%%%%%%%%%% TITLE PAGE %%%%%%%%%%%%%%%%%%%%%%%%%%%%%%%%%
%
\draft
\preprint{
   \vbox{\baselineskip=14pt%
   \rightline{FSU-HEP-971113}\break 
   \rightline{CERN-TH.97-358} }
}

\title{
NEXT-TO-LEADING ORDER SLEPTON PAIR PRODUCTION AT HADRON COLLIDERS
}

\author{
Howard Baer$^1$, B.W. Harris$^1$ and Mary Hall Reno$^{2,3}$
}

\address{
$^1$Department of Physics,
Florida State University,
Tallahassee, FL 32306-4350, USA
}

\address{
\hfill $^2$Department of Physics and Astronomy,
University of Iowa, 
Iowa City, IA 52242, USA
}

\address{
$^3$Theory Division,
CERN, CH-1211 Geneva 23, Switzerland
}

\date{December 1997}
\maketitle

\begin{abstract}
We compute total cross sections for various slepton
pair production reactions $\tell_L\tnu_L$, $\tell_L\bar{\tell_L}$,
$\tell_R\bar{\tell_R}$ and $\tnu_L\bar{\tnu}_L$ in next-to-leading
order QCD. For $p\bar p$ collisions at $\sqrt{S}=2$ TeV,
we find leading order cross sections to be enhanced by typically
35\% to 40\%.  For $pp$ collisions at $\sqrt{S}=14$ TeV, the enhancement
ranges from 25\% to 35\% depending on the mass of the sleptons.
We comment upon the phenomenological implications of these results.
\end{abstract}

\medskip
%{\tt$\backslash$\string pacs\{\}}
%
%\narrowtext
%
%%%%%%%%%%%%%%%%%% MAIN TEXT %%%%%%%%%%%%%%%%%%%%%%%%%%%%%%%%%%%%%%%%%%%%%%%
%
\section{Introduction}

Particle physics models that incorporate weak scale supersymmetry (SUSY)
are certainly among the most compelling possibilities for
physics beyond the Standard Model (SM)\cite{susy}. 
Numerous searches for the sparticles
of supersymmetric models have been made at both lepton and hadron 
colliders. The lack of any convincing supersymmetric signal beyond 
Standard Model backgrounds has led to a variety of limits on sparticle
masses\cite{pdg}. 

The best search limits on strongly interacting superpartners, the 
squarks $\tq$ and gluinos $\tg$, have been obtained from the
CDF and $\D0$ experiments at the Fermilab Tevatron $p\bar p$ collider.
By searching for multijet$+\eslt$ events, 
limits of $m_{\tg}\agt 144-154$ GeV (for $m_{\tq}>>m_{\tg}$)
and $m_{\tq}>212-224$ GeV (for $m_{\tq}\simeq m_{\tg}$) have been 
found\cite{cdf,dzero}. Recently, a next-to-leading order
computation of $\tg\tg$, $\tg\tq$ and $\tq\tq$ production has been 
completed\cite{spira}. As a result of this calculation, 
the theoretical uncertainty
due to choice of renormalization and factorization scales has been
reduced.  In addition, for a scale choice of the order of the 
average mass of the produced sparticles, the pair 
production cross sections were found to increase by a factor of 1 to 2  
depending upon the subprocesses and collider. 
This has led to 10-30 GeV improvements 
in the mass limits coming from the CDF and $\D0$ collaborations.

For weakly interacting sparticles such as the charginos $\tw_1$ and
sleptons $\tell_R$, $\tell_L$ and $\tnu_{\ell\,L}$ ($\ell =e,\ \mu$ or $\tau$),
the best search limits come from LEP2 experiments. Currently,
$m_{\tw_1}>85.5$ GeV for a gaugino-like chargino\cite{lep2chino}. For charged
sleptons, $m_{\te_R}\agt 75$ GeV, $m_{\tmu_R}\agt 59$ GeV and 
$m_{\ttau_R}\agt 53$ GeV (where the exact bounds depend on the lightest 
neutralino mass $m_{\tz_1}$)\cite{lep2slep}.
In addition, sneutrinos are required to have mass $m_{\tnu}\agt 41.8$ GeV
from measurements of the invisible $Z$ width\cite{l3} (three degenerate
generations of invisibly decaying sneutrinos have been assumed).
Search limits from the Tevatron collider on charginos and neutralinos
can be obtained by looking for clean trilepton events from
$\tw_1\tz_2\to 3\ell+\eslt$ production. Expected event rates are very model
dependent, so that limits are best placed as contours in parameter
space of specific models \cite{trilep}.

The possibility of searching for slepton production at 
hadron colliders has been studied in Refs\cite{aguila,bcpt}. 
In gravity-mediated models of supersymmetry breaking (SUGRA models) 
at Tevatron conditions, direct slepton pair
production occurs at low rates, although 
it may be possible to extract a signal 
at one of the upgrade options such as the Main Injector, which
will operate at $\sqrt{S}=2$ TeV and accumulate $\sim 2$ fb$^{-1}$ of
integrated luminosity, or TeV33, which will operate at the same energy,
but amass $\sim 25$ fb$^{-1}$ of integrated luminosity\cite{bcpt}.
For models where SUSY breaking is communicated by gauge 
interactions\cite{dnns}, slepton pair production can lead to final states
with two isolated photons plus two isolated leptons, 
a signature for which SM backgrounds are expected to be 
small \cite{gmpheno}.

For the CERN LHC operating at $\sqrt{S}=14$ TeV with projected data
sets of $10-100$ fb$^{-1}$ of integrated luminosity, SUGRA models 
predict the possibility of detection of directly produced slepton 
pairs \cite{bcpt,rurua}. This is likely possible in gauge-mediated
models as well. SUGRA models predict slepton masses of
$100-400$ GeV ought to be detectable; this mass range also corresponds
to the most favored range of slepton masses expected if the lightest
neutralino makes up the bulk of dark matter in the universe\cite{bbdm}.

Direct slepton pair event rates will be low
at an upgraded Tevatron, while at the LHC, slepton reconstruction will be
difficult. Theoretical improvements of the Born level 
analyses\cite{aguila,bcpt} of
slepton pair production are essential.
In this paper, we present the leading Quantum Chromodynamics (QCD) corrections 
to the direct production of slepton pairs at hadron colliders.
The calculation is described and formulae are given in Sec.\ II. 
Numerical results are shown for the Fermilab Tevatron 
and CERN LHC collider in Sec.\ III.  
The main result is that the slepton pair production 
cross sections are enhanced by 35-40\% for the Fermilab Tevatron collider, 
and 25-35\% for the CERN LHC collider relative to the leading order results.  
The theoretical uncertainties are addressed and it is found that the 
total cross section results should be accurate to $\pm 15$\%.  
Finally, we comment upon the implications of these results for slepton 
search experiments.

\section{Analytical Results}

In this section we briefly describe the calculation of QCD corrections 
to the direct production of slepton pairs at hadron colliders and then 
give the corresponding cross section formulae.  The results are very 
similar to those of the leading QCD corrections \cite{aem} to the 
SM Drell-Yan \cite{drellyan} process.  The lowest order cross sections 
have been discussed in Refs.\ \cite{deq} and \cite{bcpt}.

The colorless nature of the slepton pair final state allows one to 
factorize the production cross section into hadronic and SUSY pieces.  
The SUSY piece corresponds to the decay of a virtual SM gauge boson of 
mass $Q^2$ into a pair of slepton particles with masses $m_1$ and $m_2$. 
The decays are 
\begin{eqnarray}
W^{*} &\to& \tell_L\bar{\tnu_{\ell}}_L\, , \nonumber \\
Z^{*}\, ({\rm or}\, \gamma^{*}) &\to& \tell_L\bar{\tell_L}\, , \nonumber \\
Z^{*}\, ({\rm or}\, \gamma^{*}) &\to& \tell_R\bar{\tell_R}\, ,\ {\rm and} 
\nonumber \\
Z^{*} &\to& \tnu_L\bar{\tnu}_L\, .
\end{eqnarray}
Using the gauge invariance of the hadronic piece and the slepton 
Feynman rules, from Ref.\ \cite{hk} for example, the two body final state 
phase space integrals in the SUSY piece may be performed analytically.  
The result is proportional to $\beta^3$ where
\begin{equation}
\beta=\Biggl[1-{2(m_1^2+m_2^2)\over Q^2}+{(m_1^2-m_2^2)^2\over Q^4} 
\Biggr]^{1/2}.
\end{equation}

Having calculated the SUSY piece, the QCD corrections to the 
hadronic piece are standard and correspond to the production of a 
virtual SM gauge boson.  Gluon-bremsstrahlung is added 
to the leading order processes, giving the subprocesses
\begin{eqnarray}
\label{ann}
q\bar{q}' &\to& W^{*}\, g,\, {\rm and} \nonumber \\
q\bar{q} &\to& Z^{*}\, ({\rm or}\, \gamma^{*})\, g.
\end{eqnarray}
Singularities corresponding to 
soft gluon emission cancel when the ultra-violet renormalized virtual 
diagram contributions are added.
The remaining singularities are initial state collinear in origin and are 
mass factorized into scale dependent parton distribution functions.  We 
use the modified minimal subtraction ($\overline{\rm MS}$) scheme throughout.

At this order, in addition to the annihilation processes (\ref{ann}) 
there are also contributions from Compton scattering diagrams 
that may be obtained from those of (\ref{ann}) by crossing.  
There are no soft singularities for these processes, only initial state 
collinear singularities.

The total production cross section is given by a 
convolution of scale dependent parton distribution functions 
and hard scattering partonic subprocess cross sections. 
\begin{eqnarray}
\label{cross}
& & \sigma = \int^{S}_{(m_1+m_2)^2} dQ^2 
\int^1_{\frac{Q^2}{S}} dx_A \int^1_{\frac{Q^2}{Sx_A}} dx_B \Biggl\{
\sum_{ij=q,\bar{q}}\ 
f_{i/A}(x_A,\mu_f) f_{j/B}(x_B,\mu_f)
{d\hat{\sigma}_{qq}
\over d Q^2}
 \\ \nonumber \\ \nonumber
& & \quad +\sum_{i=q,\bar{q}}\Bigl[ f_{i/A}(x_A,\mu_f) f_{g/B}(x_B,\mu_f)
+f_{g/A}(x_A,\mu_f )f_{i/B}(x_B,\mu_f)\Bigr] {d\hat{\sigma}_{qg}
\over d Q^2}\Biggr\} , \\ \nonumber \\ \nonumber
\end{eqnarray}
where $f_{i/A}(x,\mu_f )$ are the parton distribution functions
for parton $i$ in hadron $A$ with momentum fraction $x$ evaluated
at the factorization scale $\mu_f$.  $S$ is the hadron-hadron center 
of mass energy squared which is related to $\hat{s}$, 
the parton-parton center of
mass energy squared, via $\hat{s}=x_A x_B S$.  Defining $z=Q^2/\hat{s}$ 
the hard scattering partonic subprocess cross sections are given by
\begin{eqnarray}
\label{qqcross}
& & {d\hat{\sigma}_{qq}\over dQ^2}
= \sigma_0\Biggl\{ \delta (1-z)+{\alpha_s(\mu_r)\over 2\pi} {4\over 3} \Biggl[
4(1+z^2)\Biggl({\ln (1-z)\over 1-z}\Biggr)_+ 
\\ \nonumber \\ \nonumber
& & \quad\quad -2{(1+z^2)\over 1-z}\ln z
+\Bigl({2\pi^2\over 3}-8\Bigr)\delta (1-z)+{3\over 2}P_{qq}(z)\ln 
{Q^2\over \mu_f^2} \Biggr] \Biggr\}
\\ \nonumber \\ \nonumber
\end{eqnarray}
and
\begin{equation}
\label{qgcross}
{d\hat{\sigma}_{qg}\over dQ^2}
= \sigma_0 {\alpha_s(\mu_r)\over 2\pi}{1\over 2}\Biggl[ {3\over 2}+z
-{3\over 2} z^2 +{2}P_{qq}(z)\Biggl( \ln{(1-z)^2\over z}-1+\ln 
{Q^2\over \mu_f^2} \Biggr) \Biggr] \,
\end{equation}
where $\sigma_0$ are given below, $\mu_r$ is the renormalization scale, 
and $P_{qq}$ and $P_{qg}$ are the splitting kernels \cite{ap} given by 
\begin{eqnarray}
& & P_{qq}(z) = {4\over 3}\Biggl[ {1+z^2\over (1-z)_+} +{3\over 2}\delta(1-z)
\Biggr]\\ \nonumber \\ \nonumber
& & P_{qg}(z) = {1\over 2}\Bigl[z^2+(1-z)^2\Bigr] \, .
\\ \nonumber
\end{eqnarray}
For the W exchange process,
\begin{equation}
\sigma_0 = {g^4\over 576 \pi}{\beta^3\over 2} z |D_W(Q^2)|^2
\end{equation}
with 
\begin{equation}
D_X(Q^2)={1\over Q^2-M_X^2+i\Gamma_X M_X}\ .
\end{equation}
For the $\gamma-Z$ exchange processes,
\begin{eqnarray}
& & \sigma_0 = {g^4\over 576\pi}{\beta^3\over 2} z 
{1\over 4\cos^4\theta_W}{\cal F}_{\gamma Z}(Q^2)
\\ \nonumber
\end{eqnarray}
with 
\begin{eqnarray}
& & {\cal F}_{\gamma Z}(Q^2)=
 | D_Z(Q^2)|^2
(L_q^2+R_q^2)c_\phi^2+ 
{32\cos^4\theta_W e^4e_\phi^2 e_q^2\over g^4}{1\over Q^4}
\\ \nonumber \\ \nonumber & & \quad\quad
+2(L_q+R_q)c_\phi{4\cos^2\theta_W e^2 e_\phi e_q\over g^2}{Q^2-M_Z^2\over
Q^2}| D_Z(Q^2)|^2\ .
\end{eqnarray}
The constants in ${\cal F}_{\gamma Z}$ are given by 
\begin{eqnarray}
& & T^f_3=\pm 1/2\ ({\rm weak\ isospin}),\\ \nonumber  
& & e_f=2/3,\ -1/3,\ -1,\ 0\ ({\rm electric\ charge}) ,\\ \nonumber
& & L_f = 2 T^f_3 - 2 e_f\sin^2\theta_W,\\ \nonumber
& & R_f = -2 e_f\sin^2\theta_W , \nonumber
\end{eqnarray}
and depending on whether one has a left- or right- SUSY partner,
$$c_\phi = L_f \ {\rm or}\ R_f\ ,$$
for slepton $\tilde{f}$.

\section{Numerical results and implications for detection}

Using the results described in the previous section a program was 
constructed to calculate slepton pair production cross sections according to 
Eq.\ (\ref{cross}).  For input, standard values of the electroweak parameters 
were used \cite{pdg} along with the CTEQ4M parton distribution 
functions\cite{cteq}.  A two-loop expression for $\alpha_s$ was 
used with fixed $n_f=5$.  The value of $\Lambda_{QCD}^{n_f=5}$ 
was taken from the parton distribution set.

The next-to-leading order (NLO) cross section results are
presented in Fig.\ 1 for {\it a}) $p\bar p$ collisions at
$\sqrt{S}=2$ TeV and {\it b}) $pp$ collisions at $\sqrt{S}=14$ TeV
as a function of slepton mass.  For $\tell_L\tnu_L$
production, we have assumed $m_{\tell_L}=m_{\tnu_L}$.  Then for both the 
Tevatron and the LHC, we see that $W^*\to\tell_L\tnu_L$ is the dominant 
production mechanism, while $\tell_R\bar{\tell_R}$ is the smallest.  
The $\tell_L\bar{\tell_L}$ and $\tnu_L\bar{\tnu_L}$ cross sections are 
intermediate between these two cases, and of comparable magnitude to 
each other.

The results of Fig.\ 1 were produced by setting the 
renormalization scale $\mu_r$ equal to the factorization scale 
$\mu_f$; the square of the scale was taken to be the invariant mass 
of the slepton pair $Q^2$.  To get some idea of the size of the 
uncalculated higher order terms we varied the scale from $Q/2$ to $2Q$.  
The results changed by $\pm 5$\% for the Tevatron and $\pm 2$\%
for the LHC.  The largest uncertainty comes from the parton 
distribution functions (PDFs). 
%Unfortunately, these uncertainties cannot be 
%quantified without repeating the entire global fitting procedure.  
For a crude estimate we use a set that assumes a different 
functional form for the gluon and sea quark at the initial evolution scale, 
the MRS Set R1\cite{mrs}.  This 
leads to an increase in cross section of approximately $10$\% for both 
Tevatron and LHC.  
We therefore cautiously estimate the theoretical uncertainty 
to be $\pm 15\%$.

In Fig.\ 2, we plot the ratio of next-to-leading order 
cross sections to leading order cross sections, again versus slepton mass
for the Tevatron and LHC colliders. 
The leading order results were computed using CTEQ4L parton distribution 
functions.

For the Tevatron collider, the various cross sections
have an enhancement of $36-39$\% depending upon the slepton mass.
The correction is nearly the same for all four slepton pair production
reactions, as might be expected by the form of the correction given in 
Eqns. (\ref{qqcross}) and (\ref{qgcross}).

The most promising detection strategy for
direct production of sleptons at the Tevatron in SUGRA-type models is via 
detection of acollinear dilepton pairs from, for instance, 
$p\bar{p}\to \tell_R\bar{\tell_R}\to \ell\bar{\ell}+\eslt$. Detailed 
simulations of this reaction have been performed in Ref.\ \cite{bcpt}.
The major SM background to the clean dilepton signal comes from 
$WW\to \ell\bar{\ell'}+\eslt$. Ref.\ \cite{bcpt} evaluated two case
studies for slepton pair production. After cuts, case 1, with
$m_{\tell_R}\simeq 80$ GeV, had 18 fb of signal, while case 2 with
$m_{\tell_R}\simeq 102$ GeV had 9 fb of signal. The background from $WW$
at lowest order was evaluated at 36 fb. Thus, with 2 fb$^{-1}$ of integrated
luminosity, neither signal case would be visible by itself at the $5\sigma$
level. Incorporation of
the NLO enhancement to both signal and background ($K\simeq 1.3$ 
for $WW$ production at the Tevatron\cite{ohnemus}) pushes case 1 above the 
$5\sigma$ level, and into observability. 
In general, for the Tevatron, supersymmetric 
clean dilepton pair events can also come 
from chargino/neutralino pair production, and the net signal would likely 
come from a variety of different sparticle production reactions. 
The reach for SUGRA type SUSY models in the clean dilepton channel has been
presented in Ref.\ \cite{bcpttev}

For the case of the CERN LHC collider, Fig.\ 2{\it b} shows that 
the various slepton pair cross sections
have an enhancement of $25-36$\% depending upon slepton mass.  
In Ref.\ \cite{bcpt,rurua}, it was found that a clean $\ell\bar{\ell}+\eslt$ 
signal could be seen above SM backgrounds for 
$100\alt m_{\tell_R}\alt 400$ GeV. With hard enough cuts, the signal was shown 
to come almost exclusively from slepton pair production rather than
from other SUSY particle production processes. However, direct reconstruction
of parent slepton masses would be extremely difficult. 
Possibly the best option for
a slepton mass measurement at the LHC might be from matching the total signal
rate to expected rates from simulations of slepton production and decay
for different mass sleptons. In this case, the most precise estimate of
the slepton production cross sections would be needed. 

\section{Conclusions}

The total cross sections for various slepton
pair production reactions $\tell_L\tnu_L$, $\tell_L\bar{\tell_L}$,
$\tell_R\bar{\tell_R}$ and $\tnu_L\bar{\tnu}_L$ were computed in 
next-to-leading order QCD.  For $p\bar p$ collisions at $\sqrt{S}=2$ TeV,
leading order cross sections were enhanced by typically
35\% to 40\%.  For $pp$ collisions at $\sqrt{S}=14$ TeV, the enhancement
ranges from 25\% to 35\% depending on the mass of the sleptons.  
The theoretical uncertainty resulting from variations in the scale and parton 
distributions was found to be approximately $\pm 15\%$.  
The NLO enhancements of the cross sections at Tevatron energies push
some predictions for signals at the MI above the 5$\sigma$ level. 
At the LHC, where a slepton mass measurement may be made indirectly
on the basis of event rates rather than by reconstruction, the
NLO predictions for slepton pair production rates would
be essential.

%%%%%%%%%%%%%%%%%%%%%%%%% ACKNOWLEDGEMENTS %%%%%%%%%%%%%%%%%%%%%%%%%%%%%%%%%%%
%
%\newpage
\acknowledgments
We thank X. Tata for comments on the manuscript.
This research was supported in part by the U.~S. Department of Energy
under contract number DE-FG05-87ER40319 and National Science Foundation 
grant PHY-9507688.
%
%%%%%%%%%%%%%%%%%%%%% REFERENCES %%%%%%%%%%%%%%%%%%%%%%%%%%%%%%%%%%%%%%%%%%%%%
%

\newpage
%
%%%%%%%%%%%%%%%%%%%%%% FIGURE CAPTIONS %%%%%%%%%%%%%%%%%%%%%%%%%%%%%%%%%%%%%%
%
%
\centerline{\bf \large{Figure Captions}}
\begin{description}
\item[Fig.\ 1]
The NLO total cross section 
for pair production of a single generation of sleptons:
$\bar{\tilde{e}}_L\tilde{\nu}_L+\tilde{e}_L\bar{\tilde{\nu}}_L$,
$\bar{\tilde{e}}_L\tilde{e}_L$, $\bar{\tilde{e}}_R\tilde{e}_R$
and $\bar{\tilde{\nu}}_L\tilde{\nu}_L$, 
via the Drell-Yan 
mechanism versus slepton mass, for {\it a}) $p\bar{p}$ collisions at
$\sqrt{S}=2$~TeV and {\it b}) $pp$ collisions at $\sqrt{S}=14$ TeV.
The slepton masses are assumed to be degenerate.
We convolute with CTEQ4M parton distribution functions.
\item[Fig.\ 2]
The ratio of NLO to LO slepton pair production cross sections
versus slepton mass, for {\it a}) $p\bar{p}$ collisions at
$\sqrt{S}=2$~TeV and {\it b}) $pp$ collisions at $\sqrt{S}=14$ TeV.
The slepton masses are assumed to be degenerate.
We convolute with CTEQ4M PDFs in the NLO case and CTEQ4L PDFs in 
the LO case.
\end{description}
%
%%%%%%%%%%%%%%%%%%%%%%%%FIGURES%%%%%%%%%%%%%%%%%%%%%%%%%%%%%%%%%%%

\newpage
% Fig. 1
\begin{figure}
\centerline{\hbox{\psfig{figure=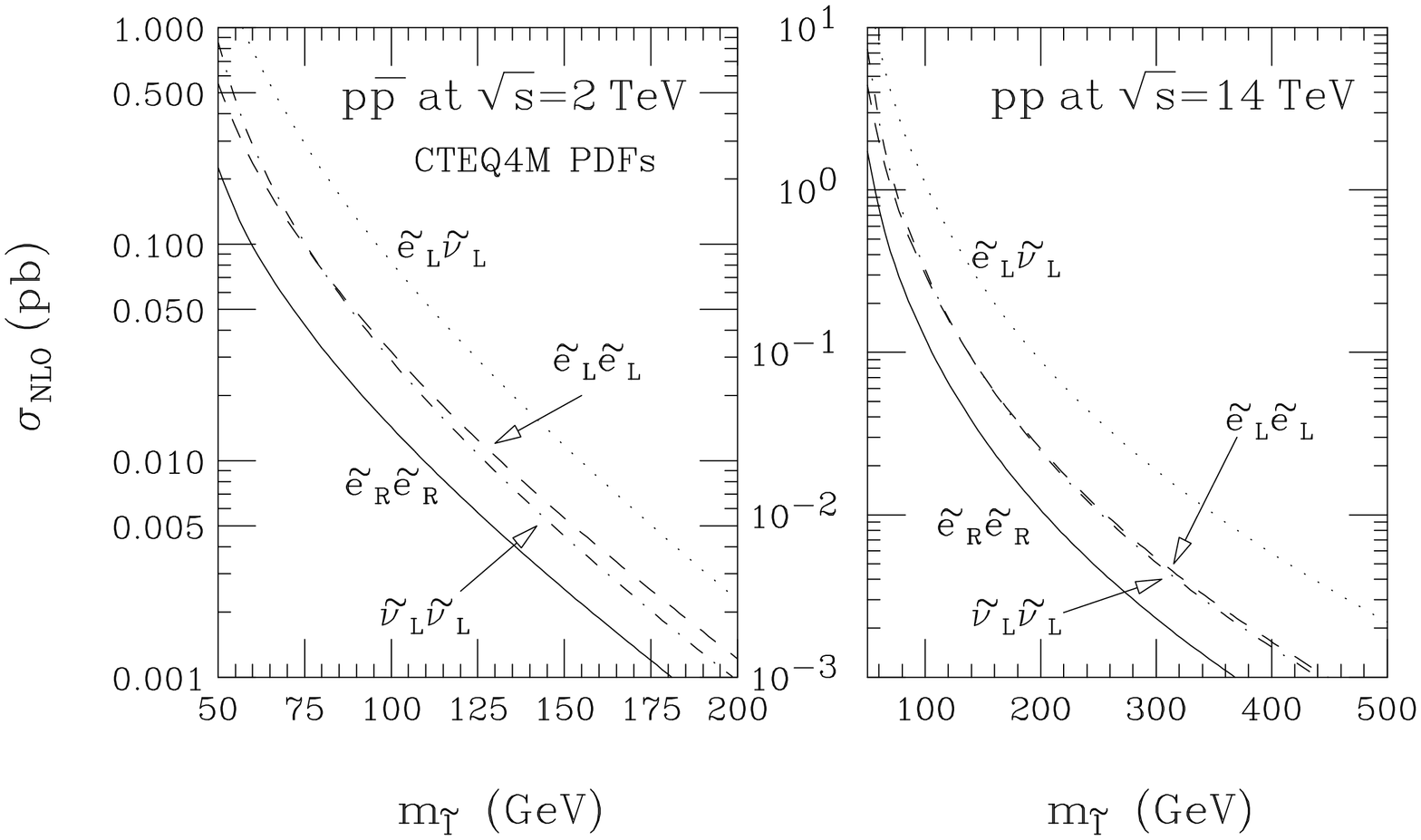,angle=90,width=6.0in,height=4.5in}}}
\caption{}
\end{figure}
% Fig. 2
\begin{figure}
\centerline{\hbox{\psfig{figure=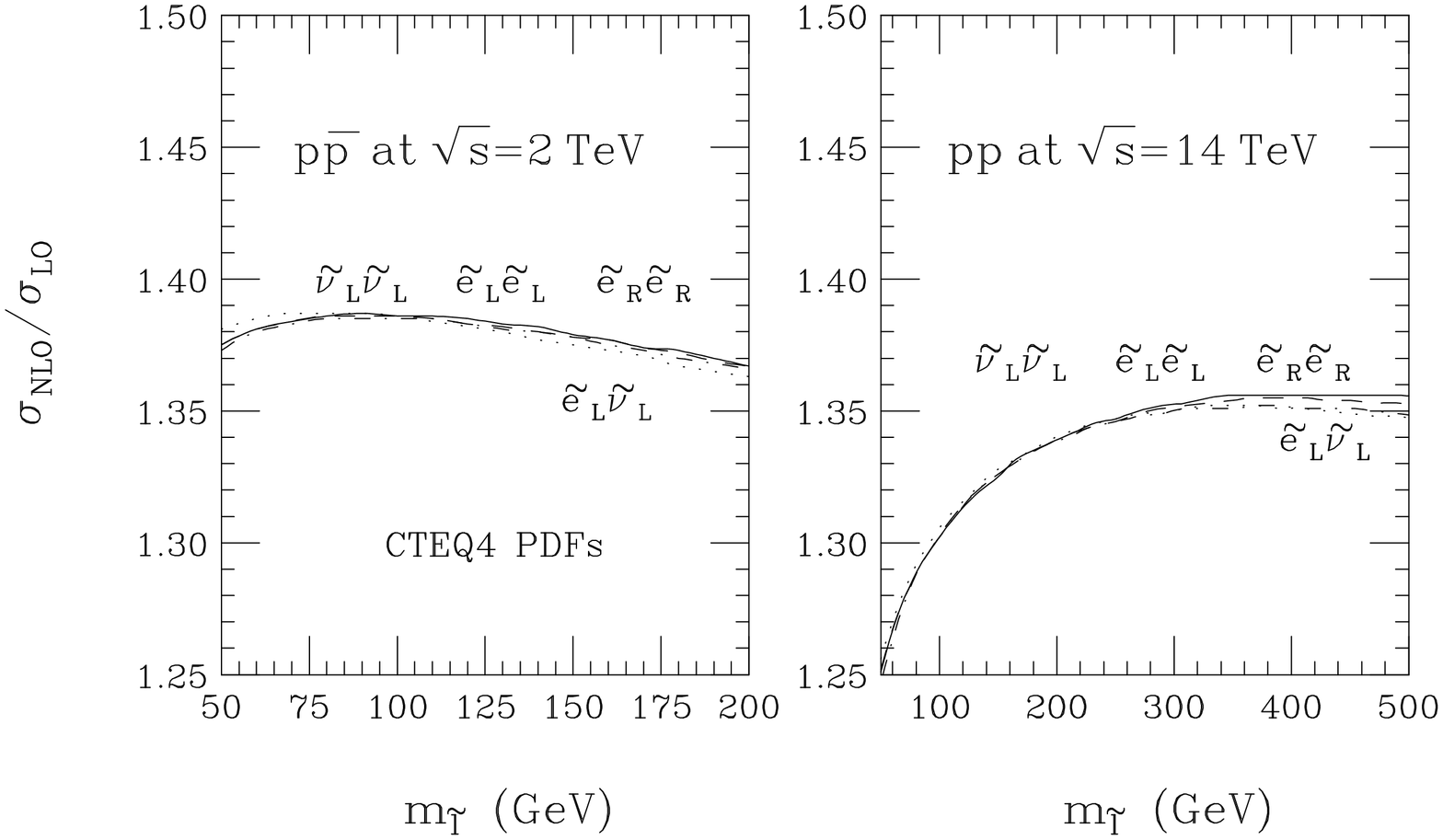,angle=90,width=6.0in,height=4.5in}}}
\caption{}
\end{figure}


\begin{references}
%
\bibitem{susy}
X. Tata, in {\it Theoretical Advanced Study Institute in Elementary 
Particle Physics (TASI '95), Colorado, 1995}, edited by D.E. Soper 
(World Scientific Press, Singapore, 1996), p.163;
M. Drees, APCTP-05, [hep-ph/9611409] (1996); 
S. Martin, [hep-ph/9709356] (1997);
for a phenomenological review, see
H. Baer {\it et al.},
in {\em Electroweak Symmetry Breaking and New Physics at the TeV Scale},
edited by T. Barklow, S. Dawson, H. Haber, and J. Seigrist, 
(World Scientific Press, Singapore, 1996), p. 216.
%
\bibitem{pdg} See R.M. Barnett {\it et al.} (Particle Data Group) 
in {\it Review of Particle Physics}, Phys. Rev. {\bf D} 1-I, 1 (1996).
%
\bibitem{cdf} F. Abe {\it et al.} (CDF Collaboration), 
Phys. Rev. Lett. {\bf 76}, 2006 (1996).
%
\bibitem{dzero} Abachi {\it et al.} ($\D0$ Collaboration), 
Phys. Rev. Lett. {\bf 75}, 618 (1995).
%
\bibitem{spira} W. Beenakker, R. Hopker, M. Spira, and P. M. Zerwas,
Phys. Rev. Lett. {\bf 74}, 2905 (1995), Z. Phys. {\bf C} 69, 163 (1995), 
Nucl. Phys. {\bf B492}, 51 (1997).
%
\bibitem{lep2chino} R. Barate {\it et al.} (ALEPH Collaboration),
[hep-ex/9710012]; P. Abreu {\it et al.} (DELPHI Collaboration),
CERN-PPE-97-107 (1997);
M. Acciarri {\it et al.} (L3 Collaboration),  CERN-PPE-97-130 (1997);
K. Ackerstaff {\it et al.} (OPAL Collaboration), [hep-ex/9708018] (1997).
%
\bibitem{lep2slep} R. Barate {\it et al.} (ALEPH Collaboration),
Phys. Lett. B {\bf 407}, 377 (1997);
P. Abreu {\it et al.} (DELPHI Collaboration),
Phys. Lett. B {\bf 387}, 651 (1996);
M. Acciarri {\it et al.} (L3 Collaboration), see Ref. \cite{lep2chino};
K. Ackerstaff {\it et al.} (OPAL Collaboration), 
Phys. Lett. B {\bf 396}, 301 (1997).
%
\bibitem{l3} O. Adriani {\it et al.} (L3 Collaboration),
Phys. Rep. {\bf 236}, 1 (1993). 
%
\bibitem{trilep} F. Abe {\it et al.} (CDF Collaboration),
Phys. Rev. Lett. {\bf 76}, 4307 (1996); B. Abbott {\it et al.} 
($\D0$ Collaboration), hep-ex/9705015 (1997).
%
\bibitem{aguila} F. del Aguila and Ll. Ametller, Phys. Lett. {\bf B261},
326 (1991).
%
\bibitem{bcpt} H. Baer, C. H. Chen, F. Paige, and X. Tata, 
Phys. Rev. {\bf D} 49, 3283 (1994).
%
\bibitem{dnns} M. Dine, A. Nelson, Y. Nir, and Y. Shirman,
Phys. Rev. {\bf D} 53, 2658 (1996).
%
\bibitem{gmpheno} S. Dimopoulos, S. Thomas, and J. Wells, 
Phys. Rev. {\bf D} 54, 3283 (1996);
S. Ambrosanio, G. Kane, G. Kribs, S. Martin, and S. Mrenna,
Phys. Rev. {\bf D} 54, 5395 (1996);
H. Baer, M. Brhlik, C. H. Chen, and X. Tata, 
Phys. Rev. {\bf D} 55, 4463 (1997).
%
\bibitem{rurua} D. Denegri, W. Majerotto, and L. Rurua, 
[hep-ph/9711357] (1997).
%
\bibitem{bbdm} H. Baer and M. Brhlik, Phys. Rev. {\bf D} 53, 597 (1996).
%
\bibitem{aem}G. Altarelli, R.K. Ellis, and G. Martinelli, 
Nucl. Phys. {\bf B157}, 461 (1979).
%
\bibitem{drellyan}S. Drell and T. Yan, 
Phys. Rev. Lett. {\bf 25}, 316 (1970).
%
\bibitem{deq} S. Dawson, E. Eichten and C. Quigg, 
Phys. Rev. {\bf D} 31, 1581 (1985).
%
\bibitem{hk}H.E. Haber and G.L. Kane, Phys. Rep. {\bf 117}, 75 (1985).
%
\bibitem{ap}G. Altarelli and G. Parisi, Nucl. Phys. {\bf B126}, 298 (1977).
%
\bibitem{cteq} H. L. Lai {\it et al.} (CTEQ Collaboration),
Phys. Rev. {\bf D} 55, 1280 (1997).
%
\bibitem{mrs}A.D. Martin, R.G. Roberts, and W.J. Stirling, 
Phys. Lett. B {\bf 387}, 419 (1996).
%
\bibitem{ohnemus}J. Ohnemus, Phys. Rev. {\bf D} 44, 1403 (1991).
%
\bibitem{bcpttev} H. Baer, C. H. Chen, C. Kao, and X. Tata,
Phys. Rev. {\bf D} 52, 1565 (1995); H. Baer, C. H. Chen, F. Paige, and X. Tata,
Phys. Rev. {\bf D} 54, 5866 (1996).
%
\end{references}
\end{document}